\documentclass[10pt,twocolumn,aps,prl,superscriptaddress,reprint]{revtex4-2}

\usepackage[latin9]{inputenc}
\usepackage{setspace}

\usepackage{bm}
\usepackage{amsmath}
\usepackage{graphicx}
\usepackage[unicode=true,
 bookmarks=false,
 breaklinks=false,pdfborder={0 0 1},colorlinks=false]
 {hyperref}
\usepackage{breakurl}

\usepackage{float}
\usepackage{threeparttable} 
\usepackage{multirow}
\usepackage{lineno}

\makeatletter

\usepackage{xcolor}
\usepackage{bm}
\usepackage{url}
\usepackage{ulem}

\makeatother

\newcommand{\be}{\begin{equation}}
\newcommand{\ee}{\end{equation}}

\newcommand{\mq}{{\mathbf{q}}}
\newcommand{\mk}{{\mathbf{k}}}

\newcommand{\blue}{\textcolor{black}}

\newcommand{\EPC}{EPC}

\newcommand{\LK}{$\lambda_{\mathbf{k}}$}
\newcommand{\PTTE}{PtTe$_{2}$}
\newcommand{\RBBL}{RbPt$_{2}$Te$_{4}$}
\newcommand{\BL}{Pt$_{2}$Te$_{4}$}
\newcommand{\BZ}{Brillouin zone}
\newcommand{\EF}{$\epsilon_{\text{F}}$}
\newcommand{\TC}{$T_{\text{c}}$}
\newcommand{\LQ}{$\lambda_{\boldsymbol{q}\nu}$}
\newcommand{\WQ}{$\omega_{\boldsymbol{q}\nu}$}
\newcommand{\af}{$\alpha^{2}F(\omega)$}
\newcommand{\afdw}{$\alpha^{2}F(\omega)/\omega$}
\newcommand{\SOC}{spin-orbit coupling}
\newcommand{\pafdw}{$\alpha^{2}F(\omega,j,\hat{n})/\omega$}
\newcommand{\deltakt}{$\Delta(\mathbf{k},T)$}
\newcommand{\deltak}{$\Delta_{\mathbf{k}}$}

\begin{document}

\title{ Enhanced Superconductivity in Bilayer PtTe$_{2}$ by  Alkali-Metal Intercalations }

 \author{Danhong Wu}
 \affiliation{Siyuan Laboratory, Guangzhou Key Laboratory of Vacuum Coating Technologies and New Energy Materials, Department of Physics, Jinan University, Guangzhou 510632, China}	

 \author{Yiping Lin}
 \affiliation{Siyuan Laboratory, Guangzhou Key Laboratory of Vacuum Coating Technologies and New Energy Materials, Department of Physics, Jinan University, Guangzhou 510632, China}	

\author{Lingxiao Xiong}
\affiliation{Siyuan Laboratory, Guangzhou Key Laboratory of Vacuum Coating Technologies and New Energy Materials, Department of Physics, Jinan University, Guangzhou 510632, China}	

\author{Junjie Li}
\affiliation{Siyuan Laboratory, Guangzhou Key Laboratory of Vacuum Coating Technologies and New Energy Materials, Department of Physics, Jinan University, Guangzhou 510632, China}

\author{Tiantian Luo}
\affiliation{Siyuan Laboratory, Guangzhou Key Laboratory of Vacuum Coating Technologies and New Energy Materials, Department of Physics, Jinan University, Guangzhou 510632, China}	

\author{Deyi Chen}
\affiliation{Siyuan Laboratory, Guangzhou Key Laboratory of Vacuum Coating Technologies and New Energy Materials, Department of Physics, Jinan University, Guangzhou 510632, China}	

\author{Feipeng Zheng}
\email{fpzheng\_phy@email.jnu.edu.cn}
\affiliation{Siyuan Laboratory, Guangzhou Key Laboratory of Vacuum Coating Technologies and New Energy Materials, Department of Physics, Jinan University, Guangzhou 510632, China}

\begin{abstract}
Layered platinum tellurium (PtTe$_{2}$) were recently synthesized with controllable layer numbers down to monolayer limit. Using $ab~initio$ calculation based on anisotropic Midgal-Eliashberg formalism, we show that by  rubidium (Rb)~intercalation, weak superconductivity in bilayer \PTTE~can be significantly boosted with superconducting \TC~= 8 K in the presence of spin-orbit coupling (SOC). The intercalant on one hand mediates the interlayer coupling and serve as an electron donor, leading to large density of states at Fermi energy. On the other hand, it increases the mass-enhancement parameter with electron-phonon coupling strength comparable to that of Pt. The  potassium intercalated bilayer \PTTE~has a comparable  \TC~to the case of Rb intercalation. The relatively high \TC~with SOC combined with experimental accessible crystal structures suggest that these superconductors are promising platforms to study the novel quantum physics associated with two-dimensional superconductivity, such as  the recently proposed type-\uppercase\expandafter{\romannumeral2} Ising superconductivity. 
\end{abstract}
\maketitle

\section{Introduction}

Transition metal dichalcogenides (TMDCs) exhibit intriguing physical properties including superconductivity, charge-density wave, Dirac semimetals, among which the two-dimensional (2D) superconductivity has received growing attention, where the reduced dimensionality leads to unique behaviors compared to their bulk counterparts. One notable example is monolayer NbSe$_{2}$: (1) the enhanced charge-density wave and reduced superconductivity in going from bulk to the monolayer~\cite{Xi2015}; (2) the transition from a two-gap superconductor (bulk) to a single-gap one (monolayer)~\cite{Noat2015a,Khestanova2018}, owning to the competition between charge-density wave and superconductivity~\cite{Zheng2019a}; (3) suppressed magnetic instability by charge-density wave~\cite{zheng2017charge}; (4) Ising superconductivity, where the inversion-symmetry-broken crystal leads to large Zeeman type spin-orbit coupling (SOC), resulting in the extremely large in-plane upper critical field~\cite{Xi2015a,Xing2017}. Besides the inversion-symmetry-broken Ising superconductivity, the large critical fields were also observed in 2D centrosymmetric systems known as type-\uppercase\expandafter{\romannumeral2} Ising superconductivity, arising from multiple degenerate orbitals with spin-orbital locking~\cite{Wang2019f,Liu2020b,Falson2019}.  Clearly, realizing 2D superconductivity in layered TMDCs is  desirable to offer potential platforms for studying the novel quantum physics and  the interplay among different orders in 2D limit.

The group \uppercase\expandafter{\romannumeral8} TMDCs, MTe$_{2}$, where M=Ni, Pd, Pt, recently attach increasing attentions for the experimentally verified  type-II Dirac semimetal~\cite{Zhang2017a,Yan2017,Huang2016},  pressure induced superconductivity in their bulk phases~\cite{Qi2020,Xiao2017a}, and novel physics in their ultrathin films~\cite{Liu2020b,Zheng2019c,Zhao2018,Lin2020,Deng2019,Liu2018a,Liu2020d}. In particular, monolayer NiTe$_{2}$ was predicted to be an intrinsic superconductor, and lithium intercalation can boost the superconducting transition temperature (henceforth $T_{\text{c}}$) of bilayer NiTe$_{2}$ up to 11.3 K~\cite{Zheng2019c}. Few-layer PdTe$_{2}$ were experimentally verified to be  type-\uppercase\expandafter{\romannumeral2} Ising superconductors though with \TC $<$ 1 K \cite{Liu2020d}. Very recently, their homologues  PtTe$_{2}$ crystals were reported to be synthesized with controllable thickness down to monolayer limit~\cite{Lin2020}. Different from NiTe$_{2}$, monolayer PtTe$_{2}$ is an intrinsic semiconductor with band gap about 0.8 eV. Interlayer coupling in PtTe$_{2}$ is expected to be stronger, due to the smaller interlayer spacing in bulk PtTe$_{2}$.  Furthermore, SOC is supposed to be stronger in PtTe$_{2}$, as Pt element is much heavier than Pd and Ni. The above differences, combined with the longing for the 2D superconductor motivate us to study the possibility of the emergence of superconductivity in 2D PtTe$_{2}$ crystals.

This Letter reports an $ab~initio$ study on electron-phonon coupling (EPC)~and superconducting properties of \PTTE. We show that the weak superconductivity~in bilayer \PTTE(\BL)~can be significantly boosted by alkali-metal intercalations. In particular, rubidium (Rb) intercalation leads to the formation of a thermodynamically stable crystal with the stoichiometry of \RBBL,~where the Rb occupy all the octahedral  sites. Based on anisotropic Midgal-Eliashberg formalism, the \TC~of the~\RBBL~is computed to be 8 K with SOC, which is very high among TMDCs. The homologues KPt$_{2}$Te$_{4}$ is shown to have comparable \TC~to \RBBL. The mechanism of the remarkable boosted superconductivity and the effect of SOC are systematically analyzed. 

\section{Computational methods}
Density-functional theory and density-functional perturbation theory calculations were performed with the exchange-correlation functional of PBE~\cite{Perdew1996}  to study the crystal structures, electronic structures, EPC~of bulk, and few-layer ~\PTTE~before and after  alkali-metal intercalations~\cite{Giannozzi2009,Kresse1996}. The norm-conserving pseudopotentials of FHI98~\cite{Fuchs1999} and ONCV ~\cite{Hamann2013} were used to describe the interaction between valance and core electrons.  The Kohn-shame valance states were expanded as plane waves below 80 Rydberg. A 18$\times$18  (18$\times$18$\times$12) $\mathbf{k}$-mesh and a 6$\times$6  (6$\times$6$\times$4) $\mathbf{q}$-mesh were adopted to calculate the ground states of charge density and phonons for few-layer (bulk) systems, whereupon the electron-phonon coupling (\EPC)~matrix element $g_{mn,\nu}(\mathbf{k},\mathbf{q})$ are calculated, which quantifies the  scattering amplitude between the electronic states with wavevector $\mathbf{k}$, band index $m$ ($\mathbf{k}$, $m$), and ($\mathbf{k}$+$\mathbf{q}$, $n$) via a phonon with branch $\nu$ and wavevector $\mathbf{q}$. Then  above quantities  are interpolated to the $\mk$-grid of 120$\times$120 (60$\times$60$\times$36) and $\mq$-grid of $60\times60$ (30$\times$30$\times$20)~\cite{mostofi2008wannier90,Ponce2016}, based on which the mass-enhancement parameter [\afdw]~are computed, where \af~is the Eliashberg spectrum, defined as:
\be
 \alpha^{2}F(\omega)= \frac{1}{2} \sum_{v} \int_{\mathrm{BZ}} \frac{d \mathbf{q}}{\Omega_{\mathrm{BZ}}} \omega_{\mathbf{q}\nu} \lambda_{\mathbf{q}\nu} \delta\left(\omega-\omega_{\mathbf{q} \nu}\right),
\ee
where $\lambda_{\mathbf{q}\nu}$ are phonon-momentum-resolved EPC constant~\cite{Ponce2016}, $\Omega_{\mathrm{BZ}}$ is the volume of the 1st \BZ, and $\delta\left(\omega-\omega_{\mathbf{q} \nu}\right)$ is replaced with a gaussian function with a broadening of 0.5 meV. Using the same $\mk$- and $\mq$-grids, the temperature-dependent superconducting gaps [\deltakt] are obtained by solving the anisotropic Midgal-Eliashberg equations with the Matsubara frequencies below 0.23 eV on imaginary axis~\cite{Ponce2016,Margine2013}, followed by performing analytic continuation to the real axis with  Pad\'e functions. 

\section{results and discussions}
Monolayer \PTTE~is composed of Te-Pt-Te triatomic layers, where each Pt is octahedral-coordinated by six Te atoms. Its bulk counterpart, 1$T$-\PTTE~ is formed by the AA stacking of such monolayers along $z$ direction, separated by an interlayer spacing $d = \blue{2.57}$~\AA~\cite{Furuseth1965}, with the hexagonal lattice constants $a$ = \blue{4.01} and $c = \blue{5.24}$~\AA~\cite{Lin2020}. The optimized lattice constants are $a$ =\blue{4.08}, and $c$ = \blue{5.27}~\AA~using the FHI98 pseudopotential~\cite{Fuchs1999}, in nice agreement with the experiment. To study the SOC effect on EPC and superconducting properties, we also adopt the ONCV~pseudopotential~\cite{Hamann2013}, which yields the optimized $a = 4.10$, and $c = \blue{5.38}$~\AA, slightly larger the experimental one. Nonetheless, we found that the above two pseudopotentials yield consistent computational results (see Sec.~S1~\cite{SM} for details). In addition to the different pseudopotentials, the van der Waals (vdW) correlations~\cite{Thonhauser2007,Sabatini2012,Hamada2014} are also found to have little influence on the electronic structures (see \blue{Sec.~S1}\cite{SM}). Therefore, we mainly report in the main text, the results of the ONCV without vdW correction. It should also be mentioned that we involve the projector augmented wave type of pseudopotentials  for more efficient calculations of $ab~initio$ molecular dynamic (AIMD), and the determination of convex hull for the Rb intercalated bilayer PtTe$_{2}$, which will be discussed soon. More details of the motivation of using the above methods can be found~\cite{PP}. The interlayer distance $d = \blue{2.57}$~\AA~of the bulk \PTTE~is much smaller than that of other layered TMDCs (e.g., 2.90~\AA~for 2$H$-NbSe$_{2}$~\cite{Selte1964}, 2.89~\AA~for $T_{d}$-WTe$_{2}$~\cite{Zheng2016d}) and even smaller than its homologues NiTe$_{2}$ (2.63~\AA)~\cite{Peacock1946}, indicating the potential strong interlayer coupling. Indeed, it was reported that the dispersive bandstructure along $c^{*}$ suppresses its superconductivity in bulk \PTTE~\cite{Kim2018}. Besides, our computational results show that the electron-doped monolayer \PTTE~can exhibit overall large EPC strength ($\lambda$) and isotropic superconducting \TC~as shown in \blue{Sec.~S2}~\cite{SM}. Therefore, it is expected that the superconductivity can be emerged in \PTTE~by  electron doping and weakening the interlayer coupling. Thus, to enhance the superconductivity in layered \PTTE, the intercalation of alkali-metal atoms should be a reasonable way, as the intercalants can act as electron donors and relieve the interlayer coupling by expanding the interlayer spacing. To realize 2D superconductivity in \PTTE, we then study the alkali-metal intercalated bilayer \PTTE.

\begin{figure}
	\centering
	\includegraphics[width=76 mm]{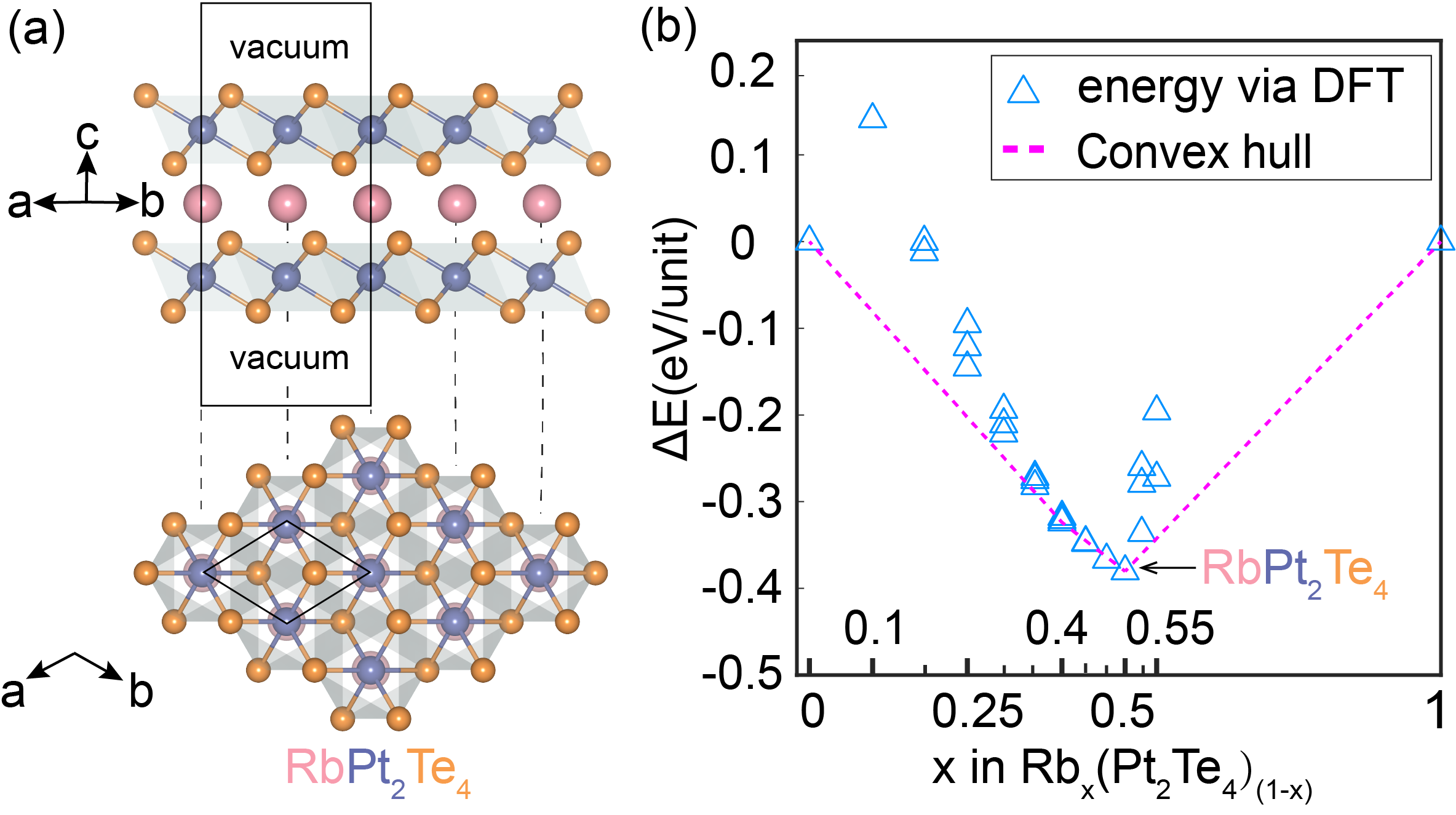} 
		\caption{  (a) Side and top views of  a slab model of rubidium intercalated  bilayer \PTTE~(schematic diagram). The black rhombus indicates a two-dimensional hexagonal unit cell. (b) Formation energies as a function of Rb concentration ($x$) calculated in a 3$\times$3 supercell via density-functional theory (DFT; see main text). The dashed pink line illustrates the convex hull between  bilayer \PTTE~($x = 0$) and the body-centered cubic rubidium crystal ($x = 1$).}
	\label{fig1}
\end{figure}

We begin by studying the energy favorable geometry structures of Rb-intercalated bilayer \PTTE, whose stoichiometry can be generated as Rb$_{m}$(Pt$_{2}$Te$_{4}$)$_{n}$, where $m$ and $n$ are integers, counting the number of Rb and Pt$_{2}$Te$_{4}$ units, respectively. We use a 3$\times$3 supercell to calculate the formation energies $\Delta E$ with respect to the Rb concentration fraction $x = m/(m+n)$, defined as  $\Delta E(x)$= $E$[Rb$_{x}$(Pt$_{2}$Te$_{4}$)$_{1-x}$] - $x $ E [Rb] - (1-$x$) $ E$[Pt$_{2}$Te$_{4}$], where $E$[Rb] and $E$[\BL] are the energies of a body-center cubic Rb crystal per atom and a bilayer \PTTE~per \BL~unit, respectively. Fig.~\ref{fig1}(b) displays the corresponding results, where $\Delta E(x)$ decrease as the increase of $x$ and reach the minimum  at $x = 0.5$, corresponding to the crystal with the stoichiometry \RBBL. Further increasing the number of Rb, the $\Delta E$ begins to increase. The resultant convex hull shown in Fig.~\ref{fig1}(b) suggests that the~\RBBL~crystal is thermodynamically stable with respect to any other stoichiometry. Fig.~\ref{fig1}(a) displays the crystal structure of the \RBBL, where the Rb atoms occupy all the octahedral  sites at the midpoints of the nearest pairs of Pt atoms in the adjacent monolayers, which is similar to the case of lithium intercalated bilayer NiTe$_{2}$~\cite{Zheng2019c}. AIMD simulations~\cite{Kresse1996,allen1989computer} further suggest that the \RBBL~crystal is stable at room temperature without structure distortion in 10 picoseconds~\blue{(see Sec. S3~\cite{SM})}.
 
\begin{figure}
	\centering
	\includegraphics[width=76 mm]{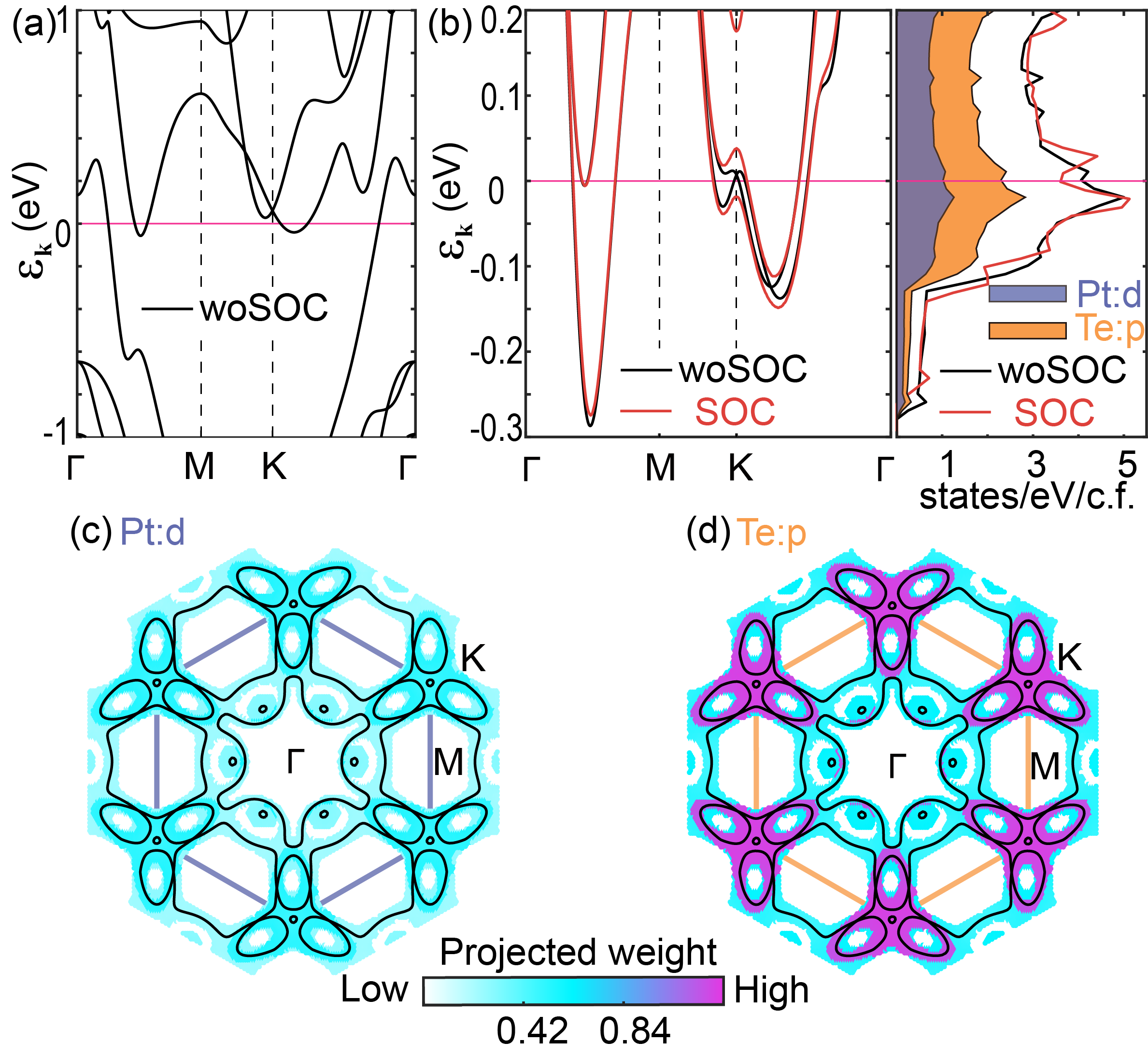} 
		\caption{  (a) Electronic bandstructure of bilayer \PTTE. (b) Electronic bandstructures of \RBBL~with/without SOC. The corresponding projected DOS (without SOC) onto $p$-like orbitals of Te and $d$-like orbitals of Pt atoms are shown in the right panel.  The projected weights (without SOC) onto the $d$-like orbitals (c) and $p$-like orbitals (d) for the electronic states of \RBBL~ within an energy window of 100 meV near \EF. The colors of the dots  represent the projection weights (see the colorbar). The black solid lines indicate the Fermi surface and the hexagons are the first \BZ.  }
	\label{fig2}
\end{figure}

Figs.~\ref{fig2}(a)~and~S4(a) display the electronic bandstructures of bilayer (henceforth \BL) and monolayer \PTTE, respectively. The monolayer is computed  to have a band gap of 0.8 eV, which agrees well with the experiment~\cite{Lin2020}. On going from the monolayer to the bilayer, a semiconductor-to-metal transition is seen with the strong modification of the electronic structure around Fermi energy (\EF). The mostly remarkable change is a band derived mainly from  the $p_{z}$-like orbitals of Te drops down and crosses \EF~(see \blue{Sec.~S4} for the projected bandstructures~\cite{SM}). This can be assigned to the overlap between the $p_{z}$-like orbitals of the Te atoms in the bilayer, suggesting the strong interlayer coupling. The calculated bandstructure of the \BL~is again consistent with the ARPES measurements~\cite{Lin2020}. After Rb intercalation, the interlayer spacing expands and the interlayer coupling weakens, accompanied by the electrons transfer from Rb to its adjacent Te layers. The synergy effects of Rb-mediated interlayer coupling and electron doping  make the bandstructure of the \RBBL~resemble that of the electron doped monolayer \PTTE, but with slight band splitting [compare Figs.~\ref{fig2}(b)~and \blue{S4(a)}]. One can see from Fig.~\ref{fig2}(b) that without SOC, there are two conduction bands crossing the \EF, leading to the formation of several electron pockets depicted by the solid black circles in Figs.~\ref{fig2}(c)--~\ref{fig2}(d). The two bands intersect at  the \BZ~corners, K and K', at the energy $\sim$\blue{12 meV} above \EF, which is responsible for the emergent tiny cirles centered at K and K'. When the SOC is included, though the Kramers degeneracy is preserved due to the inversion symmetry at the site of Rb, the double degenerated bands at K and K' are split, leading to the avoided crossing near \EF~ and a $\sim$\blue{57} meV gap opening at K and K', giving rise to a slight reduced density of states at \EF~[$N(0)$]. In addition to the small Fermi circles, the clover-shaped pockets centered at K and K' are also noted, each of them consists of three separate petals.  Centered at M and $\Gamma$ points, there are also two electron pockets. The calculated projected electronic density of states (DOS) [right panel of Fig.~\ref{fig2}(b)] and the momentum-resolved DOS for the  states near \EF~[Figs.~\ref{fig2}(c)--\ref{fig2}(d)] suggest that these states near K, K'~are significantly contributed by the $p$-like orbitals, and the remaining states are assigned to the hybridization of the $d$- and $p$-like orbitals.

\begin{figure}
	\centering
	\includegraphics[width=76 mm]{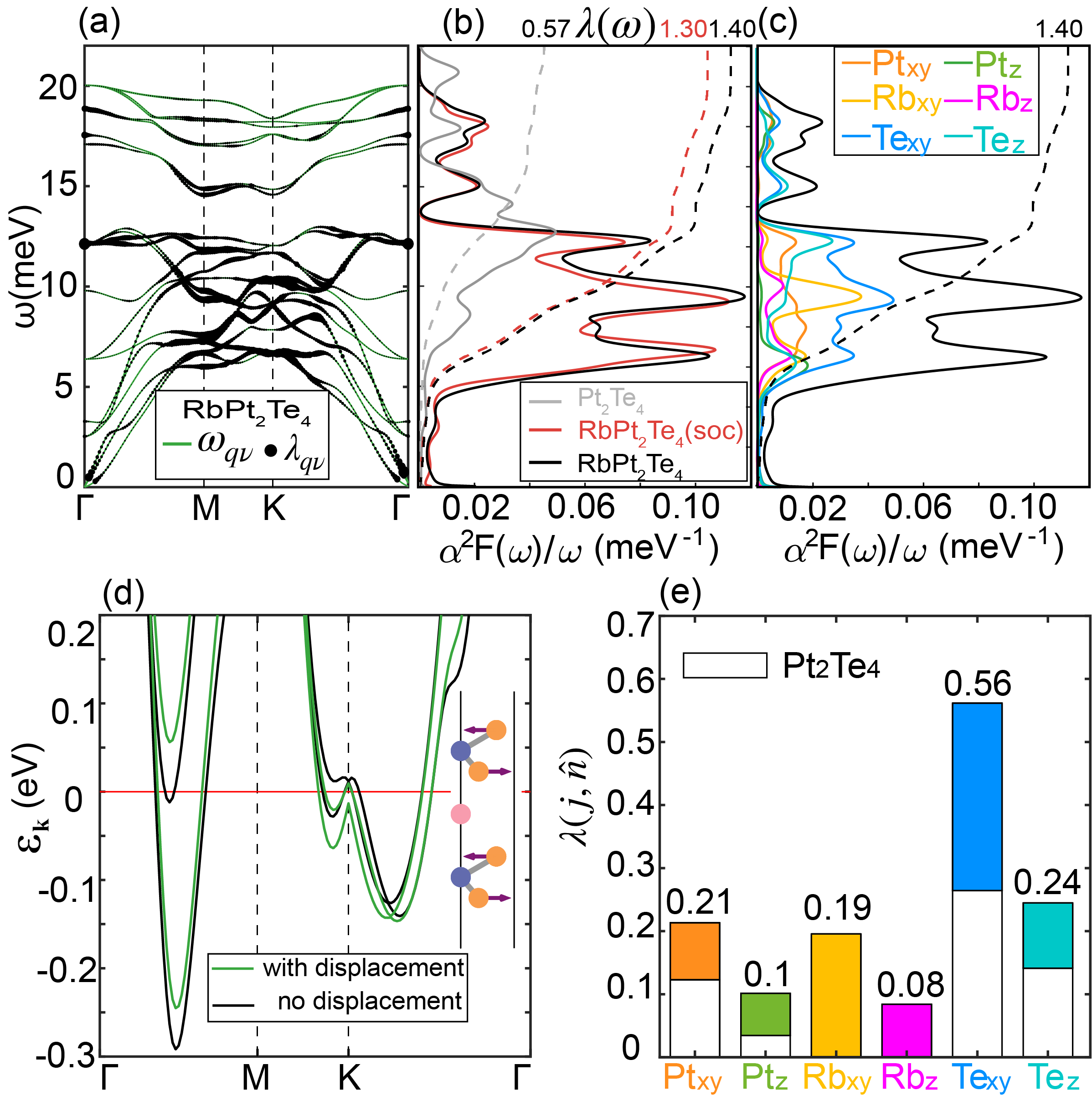} 
		\caption{ (a) Phonon dispersion (\WQ) of \RBBL~without SOC, and its corresponding \EPC~constants \LQ. (b) The mass-enhancement parameters \afdw~of  \BL~and \RBBL~with/without SOC. (c) Partial mass-enhancement parameters \pafdw, where $j$= Pt, Te, Rb, and $\hat{n}$ is the unit projection direction vector. (d) Bandstructures of \RBBL~(without SOC)~with/without atomic displacements of the phonon modes shown in the inset (see main text). The displacements are tabulated in Table S2~\cite{SM}, where one can see the Te atoms have the largest displacements of about 0.05~\AA.   (e) Partial \EPC~constants arising from various atomic vibrations: $\lambda(j,\hat{n}) = 2\int_{0}^{\infty} d\omega \alpha^{2}F(\omega,j,\hat{n})/\omega $.}
	\label{fig3}
\end{figure}

The computed phonon dispersions (\WQ) of the \RBBL~[Figs.~\ref{fig3}(a) and \blue{~S5(b)}] and Pt$_{2}$Te$_{4}$ [\blue{Fig.~S5(a)}] show that all the $\omega_{\mathbf{q}\nu}\ge0$, suggesting the dynamically stabilities of the crystals. The SOC~ is found to have little influence on \WQ~of \RBBL~[\blue{Fig.~S5(b)}], and slightly decrease the intensity of the mass-enhancement parameter [\afdw; see Fig.~\ref{fig3}(b)], but the main features of the \afdw~is similar to the case without SOC. For easier analysis, we first focus on the case without SOC. Fig.~\ref{fig3}(b) displays the comparison of the $\alpha^{2}F(\omega)/\omega$ and the corresponding accumulated EPC~strength $\lambda(\omega)=2\int_{0}^{\omega} \alpha^{2}F(\omega)/\omega d\omega$ (dashed lines) between  \RBBL~and \BL.  It is found that after the intercalation of Rb, the $\lambda(\omega)$ are remarkably enhanced in the energy region between \blue{5 -- 13} meV. Particularly, three intensive peaks are found located at \blue{6.6}, \blue{9.6} and \blue{12.3} meV, respectively. To further study the mechanism of the enhanced \EPC, we decomposed the \af~ into the contributions from in-plane ($xy$) and out-of-plane ($z$) vibrations of each atom by computing 
\begin{equation}
\alpha^{2}F(\omega,j,\hat{n})= \frac{1}{2} \sum_{v} \int_{\mathrm{BZ}} \frac{d \mathbf{q}}{\Omega_{\mathrm{BZ}}} \omega_{\mathbf{q} v} \lambda_{\mathbf{q} v} \delta\left(\omega-\omega_{\mathbf{q} \nu}\right)|
\hat{n}\cdot e^{j}_{\mathbf{q},\nu}|^{2},
\end{equation}
where  $e^{j}_{\mathbf{q},\nu}$  is the component of atom $j$ (Pt, Rb, Te) in the eigenvector of the dynamic matrix with phonon momentum $\mq$ and modes $\nu$, and $\hat{n}$ is the unit projection direction vector, which is chosen along in-plane ($xy$) and out-of-plane ($z$) directions. The computed $\alpha^{2}F(\omega,j,\hat{n})/\omega$ and the corresponding accumulated EPC constants $\lambda(j,\hat{n}) = 2\int_{0}^{\infty}d\omega\alpha^{2}F(\omega,j,\hat{n})/\omega$  are shown in Figs.~\ref{fig3}(c) and ~\ref{fig3}(e), respectively. The most remarkable change is the doubled $\lambda(\text{Te}_{xy})$ to 0.56 after the Rb intercalation [Fig.~\ref{fig3}(e)]. This is consistent with the computed $\alpha^{2}F(\omega,\text{Te}_{xy})/\omega$ [Fig.~\ref{fig3}(c)], where $\alpha^{2}F(\omega,\text{Te}_{xy})/\omega$ dominates the total spectrum of \afdw~in \blue{5 -- 13} meV, suggesting the strongly coupling of the in-plane vibration of Te atoms with the electrons.  In particular, the two pairs of double degenerated modes E$_{g}$ and E$_{u}$ can be seen at $\Gamma$ with energy of 12.08 and 12.11 meV, respectively~[Fig.~\ref{fig3}(a)], which are derived from Te$_{xy}$ vibrations [see Fig.~\blue{S6}(a)~\cite{SM}]. Among them, the two E$_{g}$ modes exhibit large \LQ. The vibrational pattern for one of the E$_{g}$ is shown in the inset of Fig.~\ref{fig3}(d). By comparing the bandstructures with and without such phonon displacements imposed on the equilibrant \RBBL~crystal, one can see that the Te$_{xy}$ displacements  have strong influences on the bands near \EF, especially around K, where a Lifshitz transition is noted [Fig.~\ref{fig3}(d)]. The secondary contributions to $\lambda$ arise from the Te$_{z}$, Pt$_{xy}$ and Rb$_{xy}$ vibrations as shown in Fig.~\ref{fig3}(e). By further examining the corresponding spectra in Fig.~\ref{fig3}(c), one can see that the contribution of Te$_{z}$ vibration is mainly  associated with the peaks at \blue{6.6} and \blue{12.3} meV, while Pt$_{xy}$ exhibits a relatively uniform contribution in the whole energy region. Interestingly, the Rb also exhibits moderate contribution to $\lambda$, as the computed $\lambda(\text{Rb}) = 0.27$ is comparable to  $\lambda(\text{Pt})$ = 0.31 [see Fig.~\ref{fig3}(e)]. The Rb vibrations promote the \afdw~peak  at \blue{9.6} meV [Fig.~\ref{fig3}(c)], being on par with the contribution of Te$_{xy}$ at the same energy.

\begin{figure}
	\centering
	\includegraphics[width=76 mm]{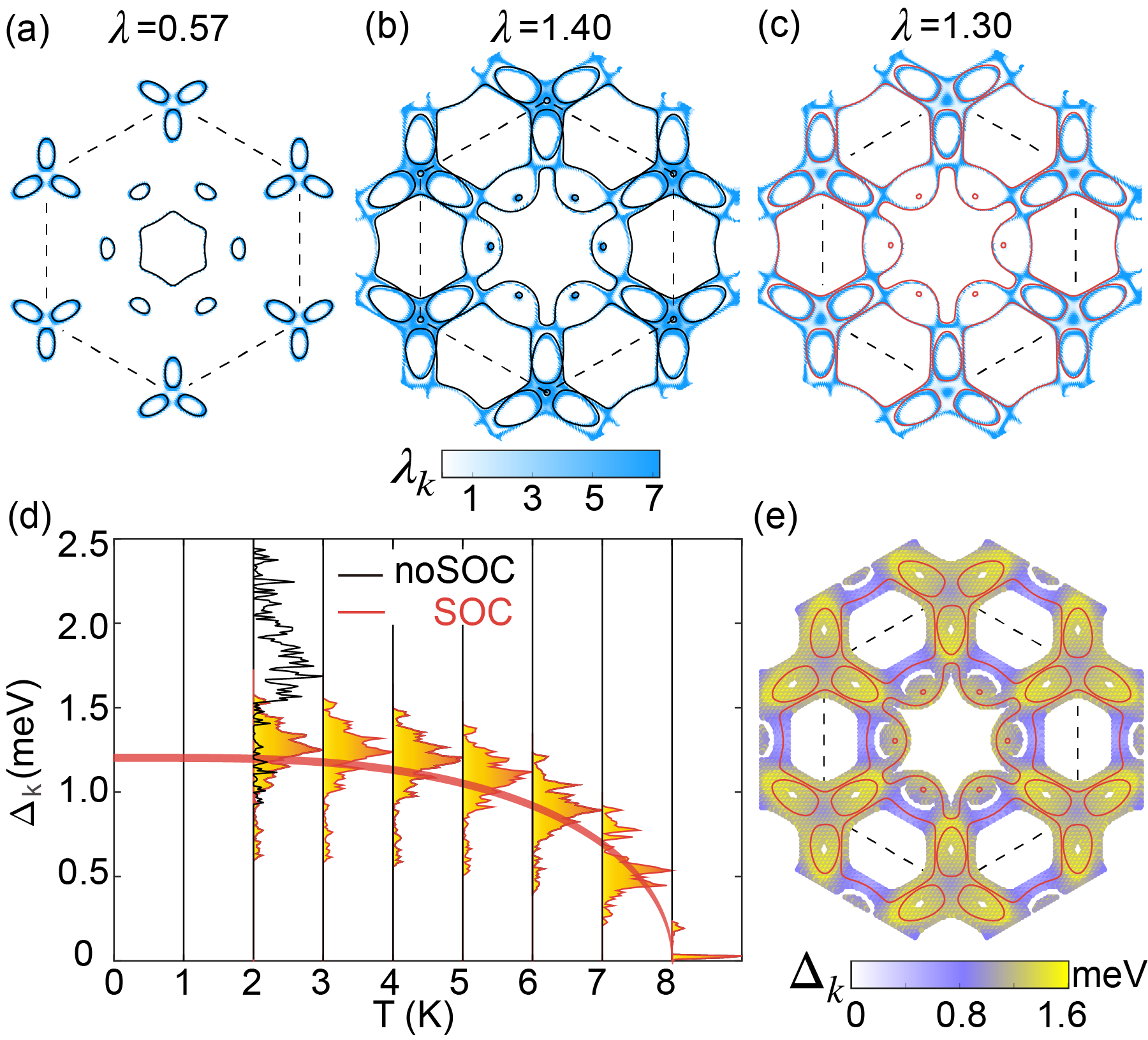} 
		\caption{ Electronic momentum-resolved \EPC~constants $\lambda_{\mathbf{k}}$ (see main text) for \BL~(a), \RBBL~without (b) and with (c)~\SOC. The corresponding Fermi surfaces are indicated by the solid lines. (d) Histograms of temperature-dependent superconducting gaps $\Delta_{\mathbf{k}}(T)$ for those electronic states $\mathbf{k}$, whose  Kohn-Sham energies are 0.1 eV around \EF. (e) The distribution of the  $\Delta_{\mathbf{k}}$ at $T = 2~$K in an extensive \BZ. SOC~is taken into account.  }
	\label{fig4}
\end{figure}

We turn to study the mechanism of the enhanced EPC~in \RBBL~by analyzing the electronic states involved in EPC. We computed the $\mathbf{k}$-resolved EPC~constant \LK, defined as follow in this work:
\begin{equation}
\begin{aligned}
	 \lambda_{\mathbf{k}}=\frac{1}{N(0)} \sum_{m, n, \nu} \int \frac{d \omega}{\omega} \int \frac{d \mathbf{q}}{\Omega_{\text{BZ}}}\left|g_{m n, \nu}(\mathbf{k}, \mathbf{q})\right|^{2}  \delta\left(\epsilon_{n \mathbf{k}}\right) \\ \times \delta\left(\epsilon_{m \mathbf{k}+\mathbf{q}}\right) \delta\left(\omega-\omega_{\mathbf{q} \nu}\right),
 \end{aligned}
\end{equation}
where $\Omega_{\text{BZ}}$ is the volume of the first \BZ, $g_{mn,\nu}(\mathbf{k},\mathbf{q})$ is the EPC matrix element, and the $\delta$ functions related to electron and  phonon energies are replaced by gaussian functions with broadening of 10 and 0.5 meV, respectively. The \LK~is related to total EPC constant by  $\lambda = \int \frac{d \mathbf{k}}{\Omega_{\text{BZ}}}\lambda_{\mathbf{k}}$.  By comparing \LK~of \BL~and \RBBL~shown in Figs.~\ref{fig4}(a) and ~\ref{fig4}(b), one can see the clover-shaped circles at K and K' are expanded, and  the additional Fermi circles emerge after Rb intercalation, promoting the $N(0)$ from 1.54 to 4.09 states/eV/Pt. Consequently, the  $\lambda$ in \RBBL~ is remarkably boosted to 1.4, as more pairs of electronic states can be involved in the \EPC. Furthermore, the contribution of the $\lambda$ of \RBBL~are almost arising from the $\mathbf{k}$ near K and K' as shown in Fig.~\ref{fig4}(b), which are mainly derived from the electronic states of Te atoms [Fig.~\ref{fig2}(d)]. By similar analysis, the states in the remaining sections of the Fermi surface, derived from the $p$-$d$ hybridization of Te and Pt atoms, also contribute to $\lambda$. The SOC~is found to reduce the \LK~near K, K' [Fig.~\ref{fig4}(c)], attributable to the reduced $N(0)$  arising from  band splitting at K, K'[Fig.~\ref{fig2}(b)]. Consequently, $\lambda$ decreases to  1.3 with SOC. Except that, the main feature of the computed \LK~is unchanged. Therefore, it is evident that the large \LK~in \RBBL~is mainly contributed by the pairing of electronic states of Te atoms near K, K' Fermi pockets.

The significantly enhanced $\lambda$ indicates the boost of the superconducting \TC. Indeed,  using McMillan-Allen-Dynes  approach~\cite{Giustino2017,McMillan1968,Allen1975} based on the calculated $\lambda$ = 1.30, the logarithmic average of the phonon frequencies $\omega_{\text{log}} = 104.5~$K, and $\mu^{*} = 0.205$ evaluated by $\mu^{*} \approx [ 0.26 N(0) /[1+N(0)]$~\cite{Bennemann1972}, the isotropic superconducting \TC~of \RBBL~is calculated to be 6.6~K  with SOC. In contrast, the computed \TC~for \BL~are only 1.1~K. The reliability of the computed isotropic \TC~is also examined by computing bulk \PTTE~as discussed in \blue{Sec.~S7}, where one can see the computed $\lambda$ for bulk \PTTE~is 0.35 and the corresponding \TC~is 0 K, in nice agreement with previous study that the bulk \PTTE~is non-superconducting with $\lambda = 0.33$~\cite{Kim2018}. To obtain more reliable superconducting \TC~for the systems with reduced dimensionality and anisotropic Fermi surface~\cite{Margine2013,Choi2002,Sanna2012}, we further evaluate the \TC~of \RBBL~by solving the full Midgal-Eliashberg gap equation~\cite{Margine2013,Choi2002}. Compared to the slight decrease of the aforementioned isotropic $\lambda$, the SOC has relatively noticeable influences on the  anisotropic superconducting properties.  Fig.~\ref{fig4}(d) shows the evolution of the energy distribution of superconducting gaps (\deltak) with respect to temperatures.  At $T$ = 2~K without SOC, the \deltak~are distributed in the energy range between 0.9 -- 2.4 meV with an average value of 1.6 meV, showing strong anisotropy. When SOC is involved, the overall decreased \deltak~ and the suppressed anisotropic distribution of  \deltak~ are seen, as the average value of \deltak~decreases to $\sim$1.2 meV, and the energy distribution range is reduced to 0.6 -- 1.6 meV. The corresponding  distribution of \deltak~in an extended \BZ~is shown in Fig.~\ref{fig4}(e), where one can see that the ~\deltak~ exhibit anisotropic distribution to some extents, in which the relatively  large ~\deltak~ comes from the Fermi pockets around K and K', dominated by the electronic states of Te atoms as analyzed before. As the temperature increases, the values of ~\deltak~gradually reduce and finally vanish at $T = 8~K$, suggesting the \TC~of the \RBBL~is $\sim$8 K, which is higher than the isotropic one (6.6 K) due to the anisotropic~\deltak. We also study the intercalation of other  alkali metal elements including lithium (Li), sodium(Na), potassium(K) and caesium(Cs), in which the KPt$_{2}$Te$_{4}$ is expected to have comparable \TC~to the \RBBL, followed by the \TC~of CsPt$_{2}$Te$_{4}$,  whereas the LiPt$_{2}$Te$_{4}$ and NaPt$_{2}$Te$_{4}$ are with relatively low \TC ~[see \blue{Sec.~S8} for details].

\section{conclusion}
   
In summary, we have predicted the enhanced EPC~and superconducting \TC~ in \RBBL~and understood the corresponding mechanisms from $ab~initio$ calculations. Firstly, according to the computed convex hull, phonon dispersions and $ab~initio$ molecular dynamic simulations, the intercalated Rb are energetically favorable to occupy all the  octahedral  sites in the interlayer gallery, forming the thermodynamically stable \RBBL~crystal. The \TC~of \RBBL~ is computed to be 8 K with the anisotropic superconducting gaps based on the anisotropic Midgal-Eliashberg formalism in the presence of SOC, though the pristine \BL~ have a very low \TC. Such a remarkable enhancement is assigned to the effects of the Rb intercalations from two sides. On one hand, the synergy effect of Rb-mediated interlayer coupling and electron doping lead to the significant promotion of $N(0)$, accompanied by the markedly enhanced EPC of Te phonons. On the other hand, the Rb directly contributes to the EPC by significantly increasing the intensity of \afdw~peak at 9.6 meV. The SOC reduces the \TC~and the anisotropic superconducting gaps by splitting the band degeneracy near \EF. The KPt$_{2}$Te$_{4}$ is proposed to have comparable \TC~to the \RBBL. The reliability of our predictions is supported by the consistent computational results of electronic structures of few-layer \PTTE, and the EPC and superconductivity of bulk \PTTE~with previous studies.

Considering bilayer PtTe$_{2}$ has been experimentally synthesized ~\cite{Lin2020}, and the intercalation of alkali-metal atoms therein can be experimentally realized~\cite{Sajadi2018,Fan2019,Nakata2019,Bianchi2012},  our predictions can be straightforwardly probed. Therefore, the experimental accessibility, combined with the relatively high superconducting \TC~with SOC will make these superconductors promising platforms to investigate intriguing novel quantum physics associated with 2D superconductivity. For example, as we have shown before, the \RBBL~crystal has three-fold rotational symmetries, which are preserved at high symmetry momenta of K and K', at which the double degenerated electronic bands very close to \EF~are noted without SOC, and they will be further split by SOC~[Fig.~\ref{fig2}(b)]. The KPt$_{2}$Te$_{4}$ also exhibits similar results (Fig.~S10). The above results suggest that the alkali-metal intercalated bilayer \PTTE~are potential candidates for realizing the type-\uppercase\expandafter{\romannumeral2} Ising superconductivity with relatively high superconducting \TC~\cite{Wang2019f}, which calls for further studies.

 \begin{acknowledgements} 
 This work is supported by National Natural Science Foundation of China (Grants No. 11804118), Guangdong Basic and Applied Basic Research Foundation (Grants No. 2021A1515010041), open project funding of  Guangzhou  Key Laboratory of Vacuum Coating Technologies and New Energy Materials (KFVEKFVE20200001). The Calculations were performed on  high-performance computation cluster of Jinan University, and Tianhe Supercomputer System.

 D.H. Wu, Y.P. Lin and L.X. Xiong contribute equally to this work.
 \end{acknowledgements}

\bibliographystyle{apsrev4-2}

\begin{thebibliography}{48}
\bibitem{PP} The optimized structure parameters of bulk PtTe$_{2}$  using the FHI98 (without vdW), which is a type of norm-conserving pseudopotential, are in nice agreement with experiment (see Table S1~\cite{SM}). Therefore, it is the best choice to using the FHI98 for all the calculations. However, we did not find the fully relativistic version of the FHI98, so that the SOC effect cannot be taken into account. Therefore, we turn to the ONCV pseudopotential, another type of norm-conserving pseudopotential, which has both the versions, so that the calculations with and without SOC can be performed, and the comparison of the results with and without SOC can be made. Therefore, we report the results of the ONCV (Figs. 2, 3 and 4) in the main text. Regarding the vdW correction, we find that the structure parameters optimized using ONCV with and without vdW are both reasonable  and the former ones are  closer to experiment (Table S1~\cite{SM}). On the other hand, we find that the density functional perturbation theory calculations are hard to converge when the vdW is included in the phonon calculations for the Pt$_2$Te$_{4}$ and RbPt$_2$Te$_{4}$. Therefore, the calculations of phonons, electron-phonon coupling and superconducting properties are using ONCV without vdW in the main text. We also examined the electronic structures computed using the ONCV with and without vdW and found that the vdW correlation has little influence on the electronic structure of \RBBL, especially for the DOS at Fermi energy (Sec.~1~\cite{SM}). Futhermore, we also find the consistent results of superconducting \TC~of \RBBL~using the ONCV (11 K, Fig.~S13\cite{SM}) and FHI98 (11.5 K, Fig.~S11\cite{SM}) without vdW and SOC. Therefore, we believe that the results obtained using ONCV without vdW correction, which we report in the main text, are reliable. It should also be mentioned that for more efficient calculations of the systems containing large numbers of atoms and electrons, we adopt projector augmented wave pseudopotentials with an energy truncation of 29.4 Rydberg, in cooperation with vdW correction, for the $ab~initio$ molecular dynamic simulations of \RBBL~using a 5$\times$5$\times$1 supercell (see Sec.~S3~\cite{SM}), and for determining the convex hull of Rb$_{x}$(Pt$_{2}$Te$_{4}$)$_{1-x}$ in a 3$\times$3$\times$1 supercell [see Fig. 1(b)]. This type of pseudopotential also yields reasonable structure parameters as shown in Table S1~\cite{SM}.
\end{thebibliography}
%

\end{document}